\pdfoutput=1 
\documentclass[12pt]{article}

\usepackage{inputenc}

\usepackage{amsmath,amssymb,bbm,color}
\usepackage{amsbsy}
\usepackage{fixmath}
\usepackage{euscript}
\usepackage{graphicx}
\usepackage{hyperref}
\usepackage{cite}
\numberwithin{equation}{section}

\usepackage{epsf} 

\def\Ordpr#1{${\cal O}(#1)_{prag}$}

\def\sigene{\delta}

\begin{document}

\begin{titlepage}


\begin{flushright}
\bf IFJPAN-IV-2015-3
\end{flushright}

\vspace{5mm}
\begin{center}
    {\Large\bf Lineshape of the Higgs boson\\ in future lepton colliders$^{\star}$ }
\end{center}

\vskip 10mm
\begin{center}
{\large S.\ Jadach$^a$ and R.A.\ Kycia$^b$}

\vskip 2mm
{\em $^a$The Henryk Niewodnicza\'nski Institute of Nuclear Physics,\\ 
Polish Academy of Sciences,\\
  ul.\ Radzikowskiego 152, 31-342 Krak\'ow, Poland}\\
\vspace{1mm}
{\em $^b$T. Ko\'sciuszko Cracow University of Technology, \\ 
Faculty of Physics, Mathematics and Computer Science, \\ 
ul. Warszawska 24, Krak\'ow, 31-155, Poland}\\
\end{center}

\vspace{5mm}
\begin{abstract}
\noindent
The effect of the photon emission (bremsstrahlung) in the cross section of 
the process of direct  production of the Higgs boson in the future high luminosity
electron and muon colliders is calculated.
It was found that cross section at the top of the Higgs boson resonance peak
is reduced by factor 0.347 for the electron collider and 0.548 for
the muon collider.
Machine spread of the centre of the mass energy of 4.2MeV 
(equal to the Higgs width)
would reduce peak cross section further, by factor 0.170 and 0.256
(QED and energy spread) for electron and muon beams respectively.
Possible uncertainties in the resummed QED calculations are discussed.
Numerical results for the lineshape cross section including QED and 
many values of the machine energy spread are provided.
\end{abstract}

\vspace{2mm}
\begin{description}
\item{\em Keywords:} QCD, the Higgs boson cross section, energy scan, 
beam spread, initial radiation state.
\end{description}

\vspace{15mm}
\begin{flushleft}
\bf IFJPAN-IV-2015-3
\end{flushleft}

\vspace{10mm}
\footnoterule
\noindent
{\footnotesize
$^{\star}$This work is partly supported by 
 the Polish National Science Centre grant UMO-2012/04/M/ST2/00240.
}

\end{titlepage}

\newpage
\tableofcontents

\newpage
\section{Introduction}

Observation of the Higgs particle at CERN 
\cite{Chatrchyan201230,Aad20121,Chatrchyan:2013lba} has opened 
the new era of  precise study of its properties.
Once its mass is known, one may reconsider a possibility of its direct
observation in the lepton annihilation $e^{-}e^{+}\rightarrow H$ or 
$\mu^{-}\mu^{+}\rightarrow H$,
at future electron colliders or dedicated muon collider.
Due to extremely small coupling of the Higgs boson to electron,
it seems at first sight that its direct production in the electron
collider is just hopeless.
However, in the Future Circular Collider with $e^{\pm}$ beams (FCCee)
considered at CERN
featuring very high luminosity,
this process would in principle be observable, 
provided one could eliminate copious background processes.
On the other hand, stronger coupling of Higgs to muons puts
dedicated muon collider into a definite advantage, 
provided decent luminosity and small beam energy spread are achieved.
In either case, machine energy spread and an additional smearing of the beam energy
due to QED bremsstrahlung are major points in the feasibility studies
of these projects.
This is why the present study was undertaken.
The influence of the beam energy spread on observability of the direct Higgs
observation in FCCee collider was already discussed at the $8$th 
FCC-ee Physics Workshop \cite{8th_FCC_Workshop_presentation}.
In this work we shall concentrate mainly on calculating effects
due to initial state radiation of multiple photons.
This will be done exploiting past experience in calculating 
very precisely similar QED effect for the Z boson productions
at LEP experiments, for instance in ref.~\cite{Jadach:2000ir}.
Similar analysis of the intial state QED corrections,
taking into accound the machine energy spread,
for $\mu^{-}\mu^{+}\rightarrow H$ process can be also found%
\footnote{We thank M. Greco and S. Dittmaier for bringing
  these works to our attention.}
in refs.~\cite{Dittmaier:2002nd} and \cite{Greco:2015yra}.

The paper is organized as follows:
After defining the Born cross section for the 
Higgs process production in the lepton annihilation,
we will discuss the effect of the initial state radiation (ISR) corrections.
We shall discuss theoretical uncertainties in the evaluation of this QED effect,
presenting numerical results for three different QED ISR formulas,
of the varying sophistication level.
The effect of the machine energy spread in the cross section will be
added in the discussion, presenting numerical results for several values
of such a spread.
Finally the analogous effects in planned muon colliders will be discussed
and numerical results will be presented.

\section{QED initial state radiation formulas}
\label{sec:ISR}

Born cross section for the Higgs particle production in $e^\pm$ collider
is given by  the (relativistic \cite{Peskin_Schroeder})
Breit-Wigner (B-W) formula \cite{Jadach:2000ir} 
\begin{equation}
 \sigma_{B}(s)=\frac{4\pi B_{ee}\Gamma_{H}^{2}}{(s-M_{H}^{2})^{2}+
 \Gamma_{H}^{2}M_{H}^{2}},
 \label{Breit-Wigner}
\end{equation}
where $M_{H}=125.7~$GeV and $\Gamma_{H} = 4.2MeV$ 
according to refs.~\cite{PDGHiggs1,PDGHiggs2}. 
The branching ratio of $e^{+}e^{-}\rightarrow H$ is $B_{ee}=5.3\cdot10^{-9}$. 
The electron and muon branching ratios are related by
the factor $m_{\mu}^{2} / m_{e}^{2}$, 
thus the above result can be obtained from $B_{\mu\mu}=2.19 \cdot 10^{-4}$,
see also refs.~\cite{PDGHiggs1,PDGHiggs2}.
There are variants of the B-W formulas with the $s$-dependent width,
but for the narrow resonance like Higgs they differ negligibly from
the above, see more discussion in the Appendix~A.

The initial state radiation correction to this process was calculated 
using formulas of ref.~\cite{Jadach:2000ir}
for Z boson production%
\footnote{See eq. (202) therein.}.
The entire initial-state \Ordpr{\alpha^2} formula of ref.~\cite{Jadach:2000ir}
integrated cross section reads:
\begin{equation}
  \begin{split}
 & \sigma_I(s) = \int\limits_0^1 dv\; \rho_I(v) \sigma_{B}(s(1-v)),
\\
&  \rho_I (v) =   e^{\delta_{YFS}}\; F(\gamma)\; \gamma v^{\gamma-1} 
     \left\{ d_{s} + \Delta_{H}(v) \right\},
 \end{split}
 \label{sigmaISR}
\end{equation}
where
\begin{equation}
 \begin{array}{ c|c|c }
  \mathrm{I} & d_{s} & \Delta_{H}(v) \\  
  \hline
  (a) & 1 & 0 \\
  (b) & 1+{\gamma\over 2}+A\frac{\alpha}{\pi} & v\left(-1 +{1\over 2} \right) \\   
  (c) & 1+{\gamma\over 2} +{\gamma^{2} \over 8}+A\frac{\alpha}{\pi} & v\left(-1 +{1\over 2} \right)
        +\gamma\left[ -{v\over 2} - {1+3(1-v)^2 \over 8} \ln(1-v)  \right]  
\end{array}
\label{eq:ISRabc}
\end{equation}
and
\begin{equation}
\begin{array}{c}
 \delta_{YFS}=\frac{\gamma}{4}+\frac{\alpha}{\pi}\left(-\frac{1}{2}+\frac{\pi^{3}}{3}\right), \\ \\ 
 \gamma = 2\frac{\alpha}{\pi}\left( \ln\frac{s}{m_{e}^{2}}-1 \right),\quad
 F(\gamma) = \frac{\exp(-C\gamma)}{\Gamma(1+\gamma)}.
\end{array}
\label{sigmaISR_additionalDefinitions}
\end{equation}
Here, $\alpha$ is the QED coupling constant, $m_{e}$ the electron 
mass and $C$ is the Euler-Mascheroni constant.
In case of the muon beams $m_e$ in $\gamma$ is replaced by $m_\mu$
and $B_{ee}$ by $B_{\mu\mu}$.

Zero spin nature of the Higgs boson instead of spin one of Z
counts negligibly in the QED ISR effects,
simply because deformation of the resonance curve is mainly due to
soft photons.
The constant $A$ which is responsible for the above spin difference is 
of order $\frac{\alpha}{\pi} \simeq 1/400$ without any logarithmic enhancement.
It is influencing mainly an overall normalization --
hence at the precision level we are aiming at, it can be safely set to zero%
\footnote{
Constant $A$ is also set to zero in ref.~\cite{Greco:2015yra},
while in ref.~\cite{Dittmaier:2002nd} vertex and real-soft contributions are provided,
but non-logarithmic constant $A$ is was explicitly obtained.
}

On the other hand soft photon exponentiation/resummation in eq.~(\ref{sigmaISR})
is critical and mandatory.
The formula of eq.~(\ref{sigmaISR}) comes from standard diagrammatic perturbative
QED calculations including Yennie-Frautschi-Suura (YFS)
exponentiation, see ref.~\cite{Jadach:2000ir} and was originally introduced
for the purpose of the algebraic validation
of the Monte Carlo program YFS2 of ref.~\cite{yfs2:1990}.
Later on it was discussed and used in many papers, see for instance
refs.~\cite{Jadach:2000ir,Jadach1992129,Jadach199977,Jadach1991173,Skrzypek1996289} 
and the references therein.

Three variants (for $I=(a),(b)$ or $(c)$) of the ISR formula in eq.~(\ref{sigmaISR})
correspond to the increasing sophistication (perturbative order)
of the non-soft collinear radiative corrections.
Change of the type of ISR formula will be used to estimate
uncertainty due to unknown/neglected QED higher orders.

\section{Machine energy spread}

In real accelerator experiments the beam is not monoenergetic, i.e., 
centre-of-mass energy $E=\sqrt{s}$ 
has spread $\sigene$ around the centre value 
$E_{0}=\sqrt{s_{0}}$ of the beam energy. 
The distribution of $E$ is usually 
well approximated  by the following Gaussian distribution
\begin{equation}
 G(E-E_{0};\sigene)=\frac{1}{\sigene\sqrt{2\pi}}e^{-\frac{(E-E_{0})^{2}}{2\sigene^{2}}}.
 \label{GaussianPDF}
\end{equation}
In case of no QED effects, the Born cross section (\ref{Breit-Wigner}) 
gets simply convoluted with the energy spectrum of eq.~(\ref{GaussianPDF}):
\begin{equation}
 \sigma_{B}^{conv} (E;\sigene) = 
 \int dE'\;  \sigma_{B}(E')\; G(E'-E;\sigene).
 \label{BornConvolution}
\end{equation}
Once QED bremsstrahlung is switched on, the following double
convolution provides realistic experimental cross section:
\begin{equation}
\begin{split}
\sigma_{I}^{conv} (E;\sigene) 
 &= \int dE'\;  \sigma_I(E') G(E'-E;\sigene)
\\
 &=\int dE'\; \int_{0}^{1}dv
 \frac{1}{\sigene\sqrt{2\pi}}e^{-\frac{(E'-E)^{2}}{2\sigene^{2}}} 
 \rho_I(v) \sigma_{B}(E'^{2}(1- v)),
\end{split}
 \label{ISRConvolution}
\end{equation}
for three variants, $I=(a), (b), (c)$ of  the radiative function (\ref{sigmaISR}).

Because of rapid decrease of the Gaussian distribution for large arguments, the energy integration range will be restricted
to $E-10 \sigene \leq E' \leq E+10 \sigene$
without any loss of the calculation reliability.
The numerical integrations in one and 
two dimensions 
require a little bit of care, because of strongly singular integrands.
The adaptive integration library functions 
of ROOT library \cite{Antcheva20092499} were used.
All results were also cross-checked using FOAM adaptive Monte-Carlo 
simulator/integrator of \cite{foam:1999,foam:2002,Slawinska:2010jn}%
\footnote{
 Integration errors were also taken from FOAM, as they are more reliable.}.


\section{Numerical results for electron-positron colliders}

\begin{table}[!t]
\centering
\begin{tabular}{c|c|c|c|c|c|c|c}
 E &$\sigma_{B}$ & $\sigma_{(a)}$& $\sigma_{(b)}$ &$\sigma_{(c)}$  
 & $\frac{\sigma_{(a)}}{\sigma_{B}}$ & $\frac{\sigma_{(b)}}{\sigma_{B}}$  
 & $\frac{\sigma_{(c)}}{\sigma_{B}}$ \\
 \hline
$M_H$           & 1.6413 & 0.5398 & 0.5693 & 0.5701 & 0.328 & 0.346 & 0.347 \\
$M_H+\Gamma_{H}$ & 0.3283 & 0.1846 & 0.195  & 0.1949 & 0.562 & 0.594 & 0.593 \\
$M_H-\Gamma_{H}$ & 0.3283 & 0.1073 & 0.1134 & 0.1133 & 0.326 & 0.345 & 0.345
\end{tabular}
\caption{\sf 
Born cross sections (in $fb$) of eq.~(\ref{Breit-Wigner})
and three variants of the ISR-corrected cross sections of eq.~(\ref{sigmaISR})
for three values of $\sqrt{s}=E = M_{H}= 125.7~$GeV  and $E=M_H\pm\Gamma_{H}$
in the electron-positron collider.
The integration error is below $0.0005fb $.
}
  \label{tab:ComparisioDistributionsPeak}
\end{table}

\begin{table}[!h]
\centering
\begin{tabular}{c|c|c|c|c|c|c|c}
E &$\sigma_{B}$ & $(1)$& $(2)$ & $(3)$ & $\frac{(1)}{\sigma_{B}}$ & $\frac{(2)}
{\sigma_{B}}$ & $\frac{(3)}{\sigma_{B}}$ \\
\hline
$M_H$           & 1.6413 & 0.5701 & 0.2795 & 0.1825 & 0.347 & 0.170 & 0.111 \\
$M_H+\Gamma_{H}$ & 0.3283 & 0.1950 & 0.2323 & 0.1753 & 0.594 & 0.708 & 0.534 \\
$M_H-\Gamma_{H}$ & 0.3283 & 0.1133 & 0.1865 & 0.155  & 0.345 & 0.568 & 0.472
\end{tabular}
 \caption{\sf 
    Cross section of the direct Higgs production process in the electro-positron collider
    including effect of ISR type (c) and the machine energy spread.
    No energy spread is in $\sigma_{(1)}$,
    while $\sigma_{(2)}$ and $\sigma_{(3)}$ include ISR and
    energy spread $\sigene = 4.2~MeV$ and $\sigene = 8.0~MeV$
    according to eq.~(\ref{ISRConvolution}).
    Reference Born cross section $\sigma_{B}$ is also included and the
    ratios with respect to Born are also provided.
    The integration error is below $0.0005fb $.
 }
  \label{tab:ComparisioDistributionsPeakALL}
\end{table}

\begin{figure}[!t]
  \centering
  \includegraphics[width=0.95\textwidth ]{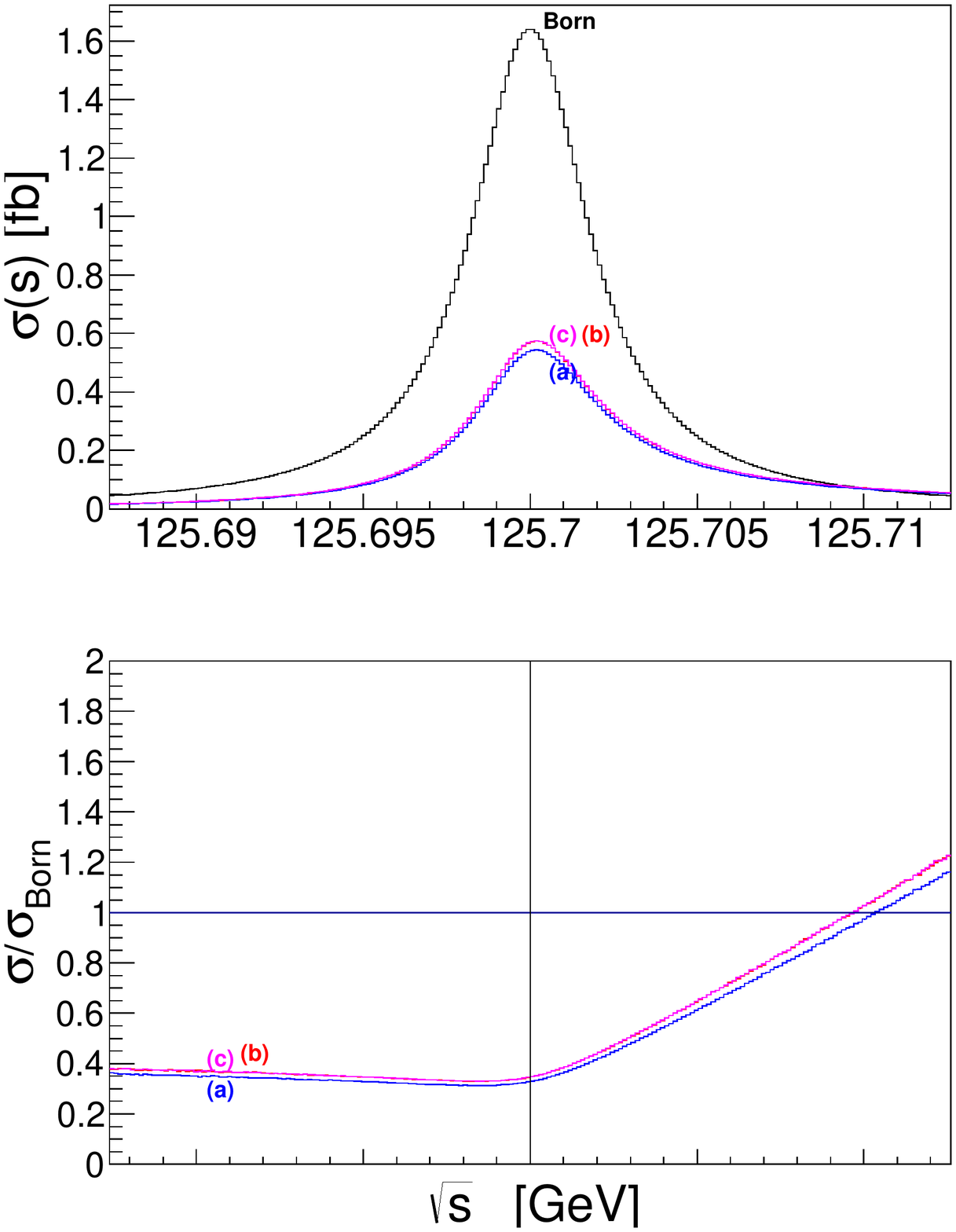}
  \caption{\sf
   Study of the pure QED effect in the Higgs line-shape for electron collider.
   The plots show Born cross sections of eq.~(\ref{Breit-Wigner}) 
   and cross section affected by QED ISR,
   following eq.~(\ref{sigmaISR}) for three types $I=a,b,c$ 
   of the QED radiator functions defined in eq.~(\ref{eq:ISRabc}).
   The ratios with respect to Born cross section are also shown.
  }
  \label{fig:sigmaISR}
\end{figure}

The Born cross sections and results from
three variants of the ISR formula of eq.~(\ref{sigmaISR})
are presented in Fig.~\ref{fig:sigmaISR}.
The case $\sigma_{(b)}$ and $\sigma_{(c)}$ are almost indistinguishable.
In addition, Tab.~\ref{tab:ComparisioDistributionsPeak}
presents the same results for the energy at the peak $E=M_H$
and near the peak $E=M_H \pm \Gamma_H$ with the 4-digit precision.
The QED uncertainty from unaccounted QED higher orders can be estimated
from Tab.~\ref{tab:ComparisioDistributionsPeak} looking
into differences between ISR type $I=(c)$ and  $I=(b)$.
It is below $0.3\%$ and is compatible with the estimate of neglected
$A$ in eq.~(\ref{sigmaISR})
(it is also comparable to the numerical integration error).
From now on we shall use ISR formula for $I=(c)$ only.

\begin{figure}[t]
  \centering
  \includegraphics[width=\textwidth ]{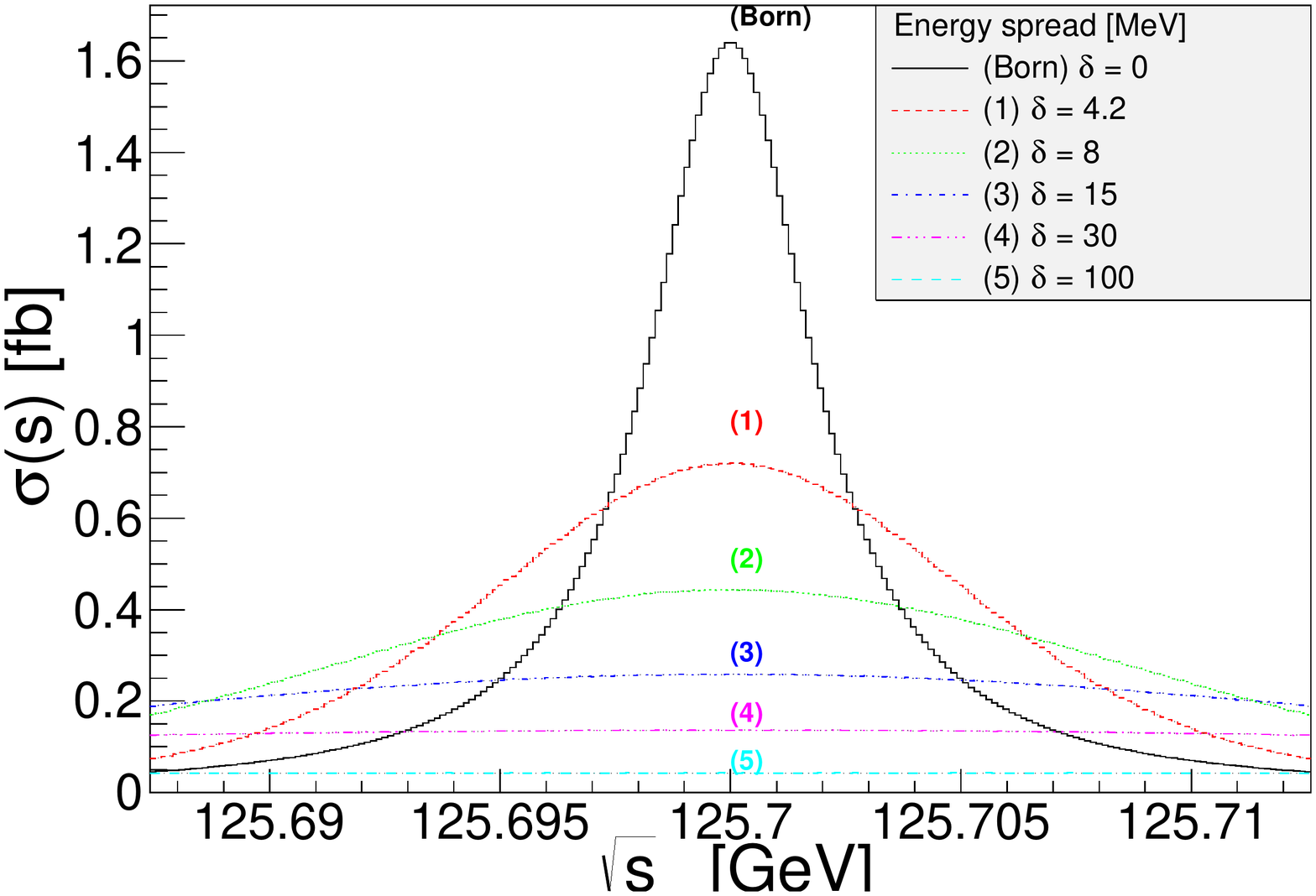}
  \caption{\sf
    Higgs production cross section in the electron-positron collider
    for several values of the machine energy spread
    $\sigene= 0,4.2,8,15,30,100$MeV.
    The QED ISR effect is not included.
  }
  \label{fig:ConvBorn}
\end{figure}

The introductory exercise on the machine energy spread is shown 
in Fig.~\ref{fig:ConvBorn} where
Higgs production cross section in the electron-positron collider
is plotted without energy spread (Born)
and for several values of the machine energy spread
$\sigene= 4.2,8,15,30,100$MeV.
The QED ISR effect is not yet included.

\begin{figure}[t]
  \centering
  \includegraphics[width=\textwidth ]{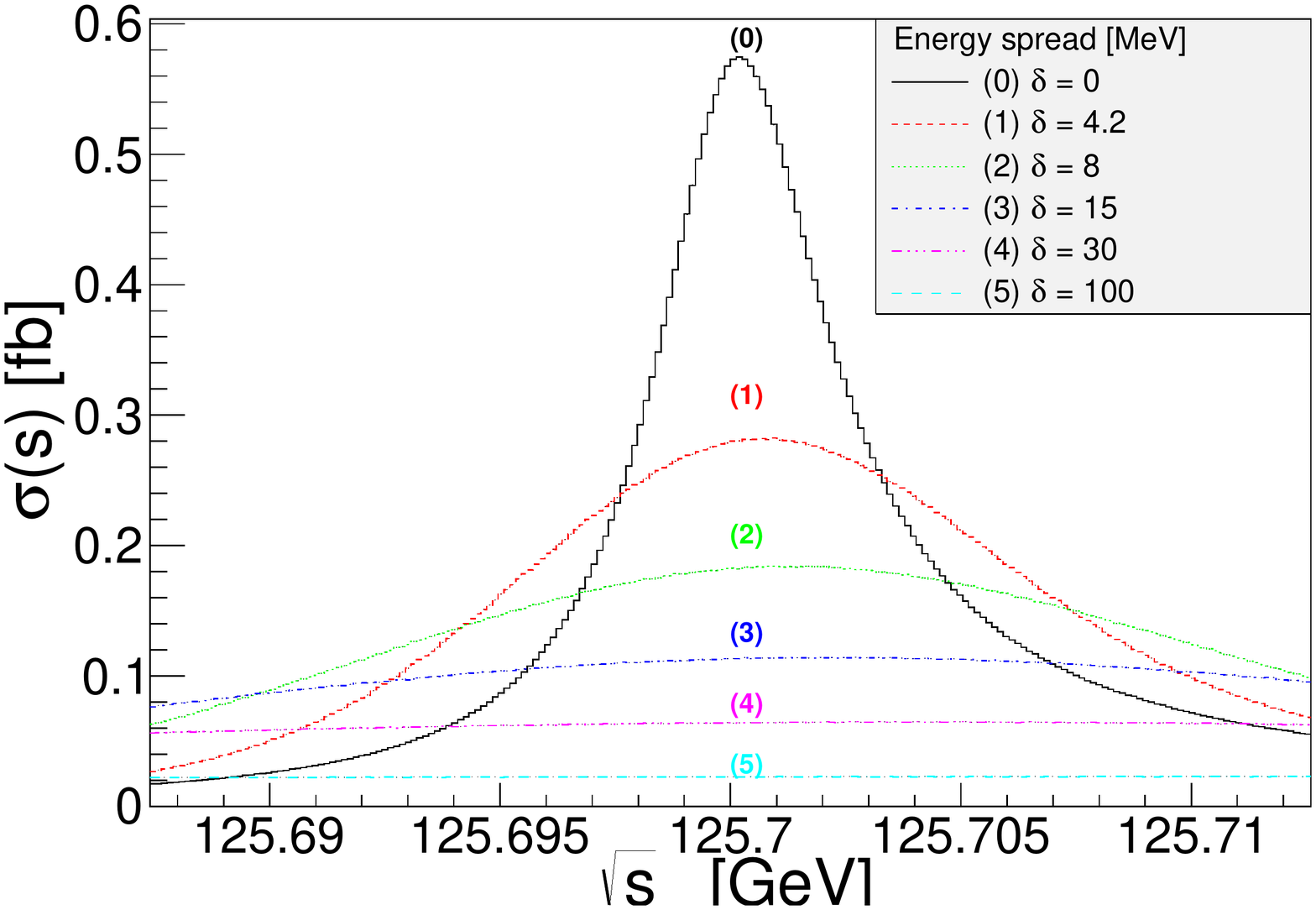}
  \caption{\sf
    Higgs production cross section in the electron-positron collider
    for several values of the machine energy spread
    $\sigene= 0,4.2,8,15,30,100$MeV.
    The QED ISR effect is included according to eq.~(\ref{ISRConvolution}),
    for ISR type $(c)$.
  }
  \label{fig:ConvOAlpha2}
\end{figure}

On the other hand Fig.~\ref{fig:ConvOAlpha2} shows our
most interesting result, that is
Higgs production cross section in the electron-positron collider including both 
QED ISR effect according to eq.~(\ref{ISRConvolution}) for ISR type $(c)$
and for several examples of the machine energy spread
parameter $\sigene= 4.2,8,15,30,100$MeV.

\begin{figure}[t]
  \centering
  \includegraphics[width=0.8\textwidth ]{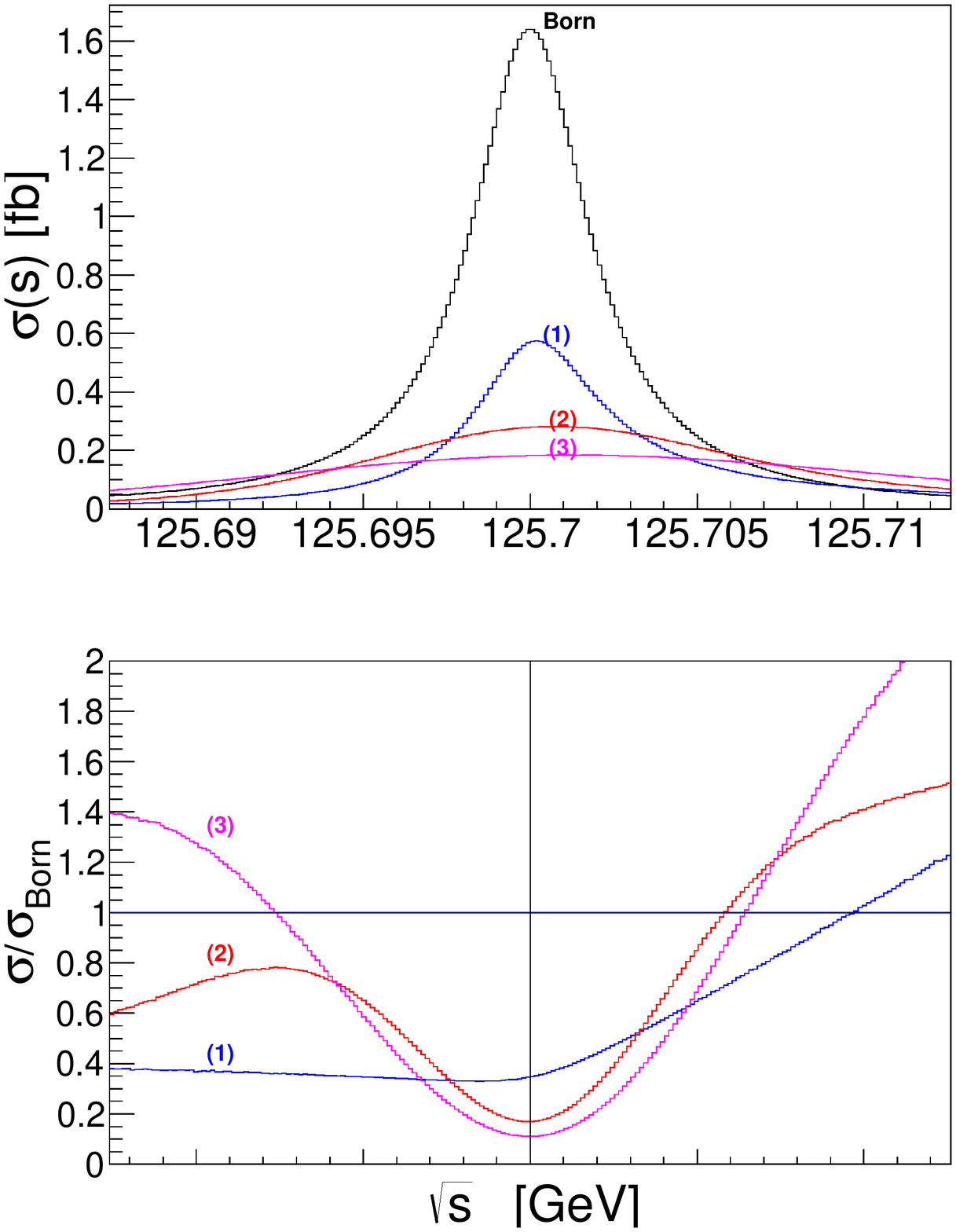}
  \caption{\sf
    Line-shape of the direct Higgs production process in the electro-positron collider
    including effect of ISR type (c) and the machine energy spread.
    No energy spread is included (only ISR is on) in the line marked with $(1)$,
    while lines marked with $(2)$ and $(3)$ include ISR and
    energy spread $\sigene = 4.2~$MeV and $\sigene = 8.0~$MeV
    according to eq.~(\ref{ISRConvolution}).
    Reference Born cross section is also shown and the
    ratios with respect to Born are also plotted.
  }
  \label{fig:Comparison}
\end{figure}

In Fig.~\ref{fig:Comparison} we plot again 
the same Higgs production cross section in the electron-positron collider 
including both  QED ISR effect and machine energy spread in the narrower
energy range and for the energy spread
parameter $\sigene= 4.2$MeV which is ambitiously aimed at the FCCee collider, 
and also for more realistic $\sigene=8.0$MeV.
The ratio of cross section with respect to Born is also shown in the lower plot there.

The same results as in Fig.~\ref{fig:Comparison} are also shown in 
Tab.~\ref{tab:ComparisioDistributionsPeakALL} for three values of the energy,
at the resonance peak $E=M_H$  and near the peak $E=M_H \pm \Gamma_H$, 
with the 4-digit precision.
As we see there, the combined reduction of the Higgs production process
at the resonance peak due to QED effect is 0.347
and goes down to 0.170 
for the machine energy spread equal to Higgs width $\sim 4.2$MeV.

The suppression factor of the peak cross section due to QED ISR
can be also quite well reproduced, and in this way cross checked,
with very simple approximate calculation presented in Appendix B.
This approximation works even better in case of the weaker
ISR effect for muon beams.

The above results can be used as an input for further studies of the practical
observability of the Higgs resonant cross section in the future
electron-positron colliders.

\begin{figure}[t]
  \centering
  \includegraphics[width=\textwidth ]{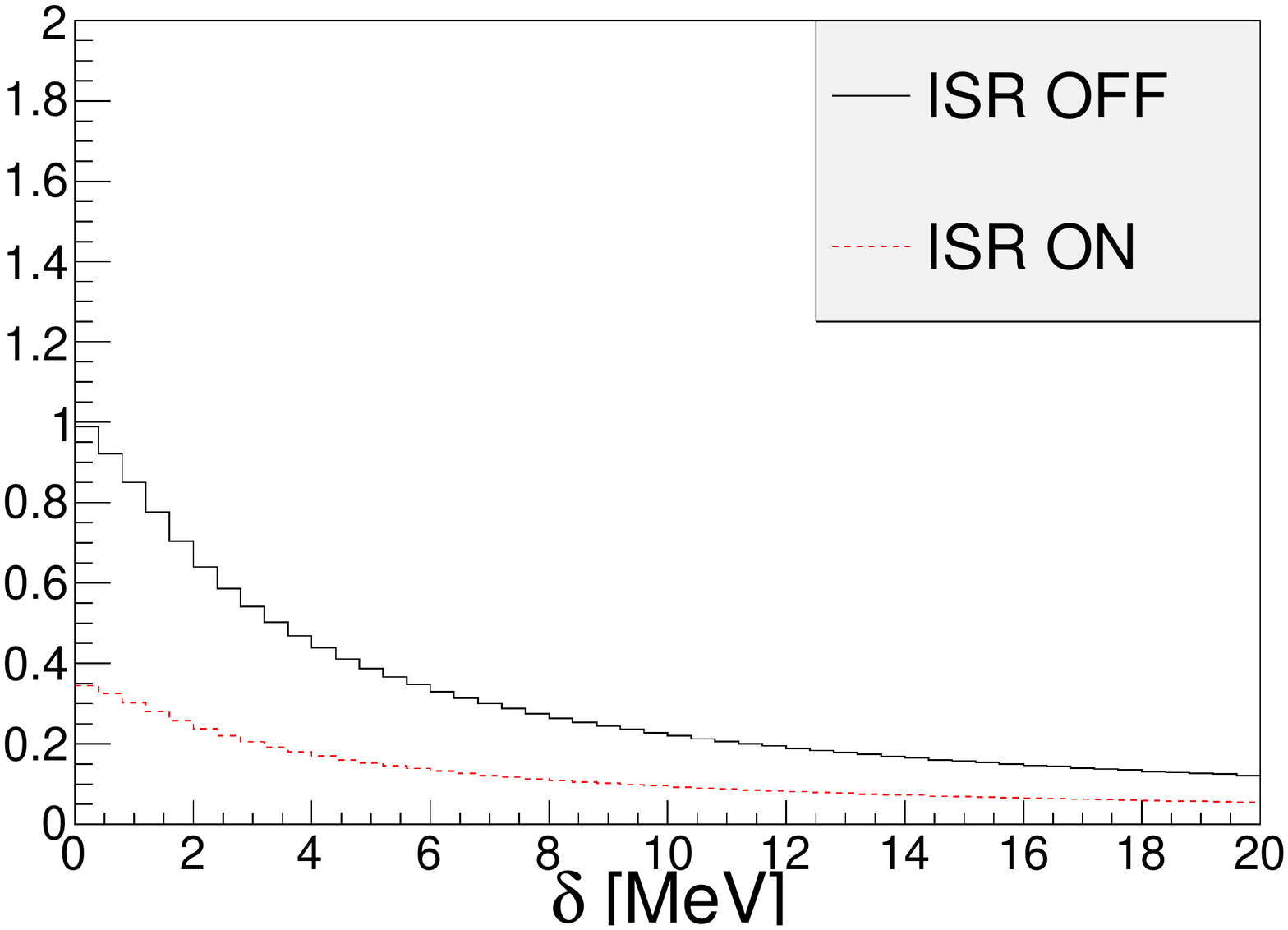}
  \caption{\sf
   The dependence of the Higgs production process 
   at the resonance peak, $E=M_H$
   in the electron-positron collider
   on the energy spread $\sigene$,
   divided by the Born cross section,
   for QED ISR switched on and off.
  }
  \label{fig:Voigt}
\end{figure}

In the above numerical exercises we have used discrete values of the machine
energy spread $\sigene$.
In Fig.~\ref{fig:Voigt} we also show the continuous dependence
on $\sigene$ of the Higgs production process at the resonance peak, $E=M_H$,
divided by the Born cross section,
both for QED ISR switched on and off.
More precisely, we are plotting there the following two ratios
\begin{equation}
  \frac{\sigma_{B}^{conv}(M_H,\sigene)}{\sigma_{B}(M_H)}
  \quad {\rm and} \quad
  \frac{\sigma_{(c)}^{conv}(M_H,\sigene)}{\sigma_{B}(M_H)},
\end{equation}
for ISR off and on respectively.

One could easily expand the above numerical exercises to more values
in the 2-dimensional space of the energy $E$ and machine energy spread $\delta$,
but we think that numerical results collected above are
complete enough
and can serve as a reliable starting point for all studies
of the Higgs resonance observability in the future electron-positron colliders.

\begin{figure}[t]
  \centering
  \includegraphics[width=0.95\textwidth ]{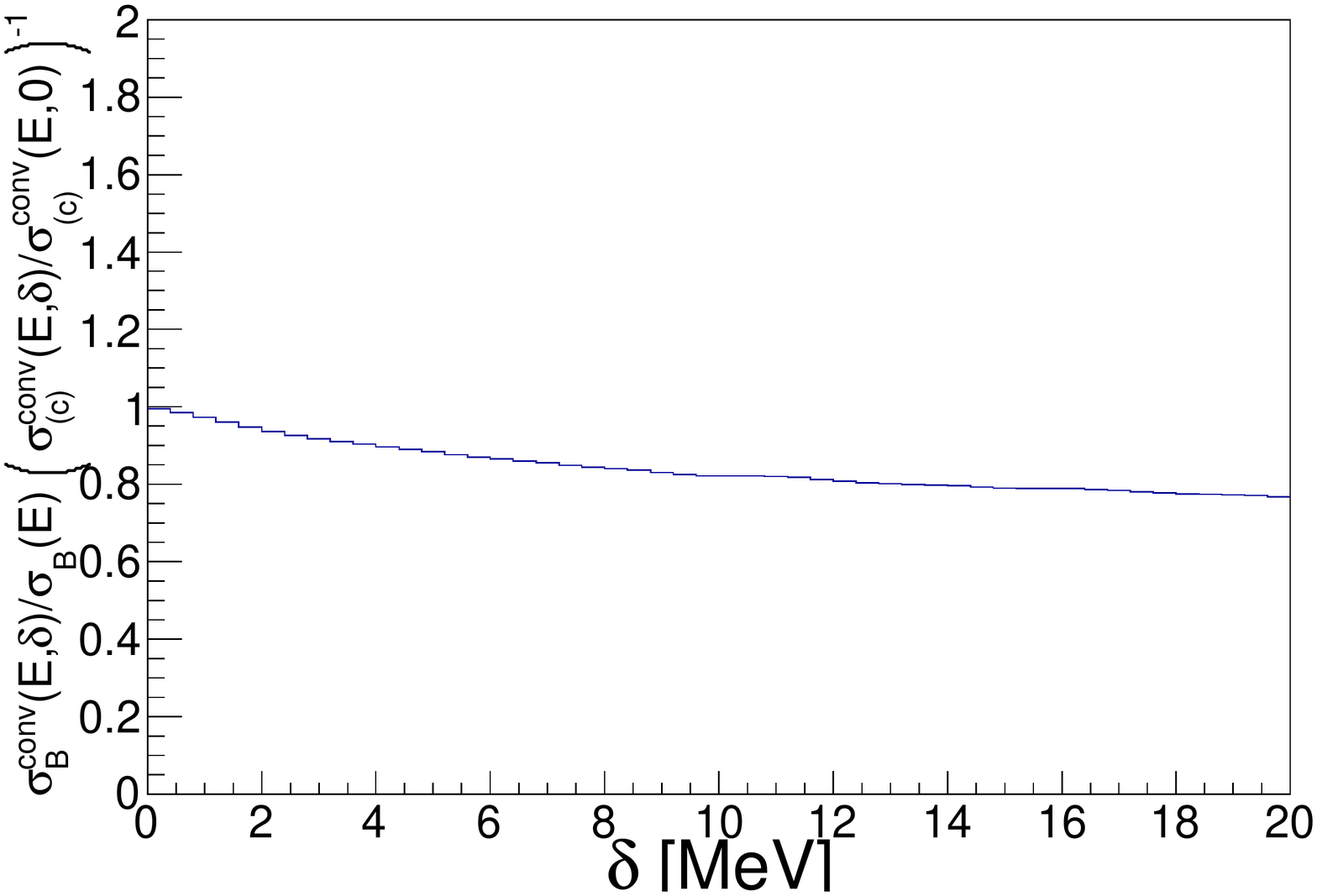}
  \caption{\sf
    Numerical cross-check of the validity of the approximation
    shown in eq.~(\ref{convolutionApproximation}) for $E=M_{H}$. 
  }
  \label{fig:voigtApproximation}
\end{figure}

It is worth to pointing out 
that because of the normalization of convolution as a function 
of $\delta$ at $E=M_{H}$ to unity we can construct the following quantity that is
close unity when $\delta/\Gamma_H$ is not too big.
In Fig. \ref{fig:voigtApproximation} the following quantity
\begin{equation}
  R_V(E,\delta)=
  \frac{\sigma_{B}^{conv}(E,\sigene)}{\sigma_{B}(E)}
  \bigg( \frac{\sigma_{(c)}^{conv}(E,\sigene)}{\sigma_{(c)}^{conv}(E,0)} \bigg)^{-1}
  \label{convolutionApproximation}
\end{equation}
is plotted.
If $R_V(E,\delta) \simeq 1$ then double convolution can be replaced
by the following handy approximation
\[
 \sigma_{(c)}^{conv}(E,\sigene) \simeq
 \sigma_{(c)}^{conv}(E,0)\;
 \frac{\sigma_{B}^{conv}(E,\sigene)}{\sigma_{B}(E)}.
\]
As seen in Fig.~\ref{fig:voigtApproximation}
the above approximation for $E=M_H$ works reasonably, 
within 10\% for $\delta \leq \Gamma_H$,
but its validity deteriorates significantly for $\delta > \Gamma_H$ 
and in such a case numerical double convolution is unavoidable.

\section{Numerical results for muon colliders}

\begin{table}[!t]
\centering
\begin{tabular}{c|c|c|c|c|c|c|c}
&$\sigma_{B}$ & $(1)$& $(2)$ & $(3)$ & $\frac{(1)}{\sigma_{B}}$ & $\frac{(2)}
{\sigma_{B}}$ & $\frac{(3)}{\sigma_{B}}$ \\
 \hline
$E_0$           & 67.8198 & 37.14 & 17.34 & 11.06 & 0.548 & 0.256 & 0.163 \\
$E_0+\Gamma_{H}$ & 13.5636 & 10.36 & 13.62 & 10.32 & 0.764 & 1.004 & 0.761 \\
$E_0-\Gamma_{H}$ & 13.5643 & 7.40 & 11.95  &  9.62 & 0.546 & 0.881 & 0.709
\end{tabular}
\caption{\sf 
  Cross section of the direct Higgs production process in the $\mu^+\mu^-$ collider
  including effect of ISR type (c) and the machine energy spread.
  The same values of the energy spread $\sigene$ an energy $E$ are used as in 
  Tab.\ref{tab:ComparisioDistributionsPeakALL}.
  The Monte Carlo integration error is below $0.03pb$.
 }
  \label{tab:muonComparisioDistributionsPeakALL}
\end{table}

The calculations of the previous section
can be easily extended to the case of muon 
collider, in the so-called Higgs factories
(for overview see, e.g., \cite{Kaplan:2014xda}),
by means of replacing $m_{e}$ with $m_{\mu}$ and 
$B_{ee}$ with $B_{\mu\mu}$ in the equations of section \ref{sec:ISR}
(see also \cite{Dittmaier:2002nd} or \cite{Greco:2015yra}). 
This kind of colliders will produce clean sample of Higgs boson
without much background and therefore 
would allow to measure Higgs mass and its properties very precisely. 
Our analysis is vital in view of 
MAP (Muon Accelerator Program) \cite{Kaplan:2014xda} at Fermilab 
and the new ideas of producing
muon monochromatic beams, that would allow
to limit the major obstacle, the machine energy spread.

Thanks to much higher branching ratio for the Higgs to muon pair,
the production cross section for the
$\mu^+\mu^- \rightarrow H$ is much higher.
Also, QED effects for muon collider are roughly factor 2 weaker,
simply because for muon beams at the Higgs peek $\gamma=0.0611$,
as compared to $\gamma=0.1106$ for electron beams.

\begin{figure}[t]
  \centering
  \includegraphics[width=0.8\textwidth ]{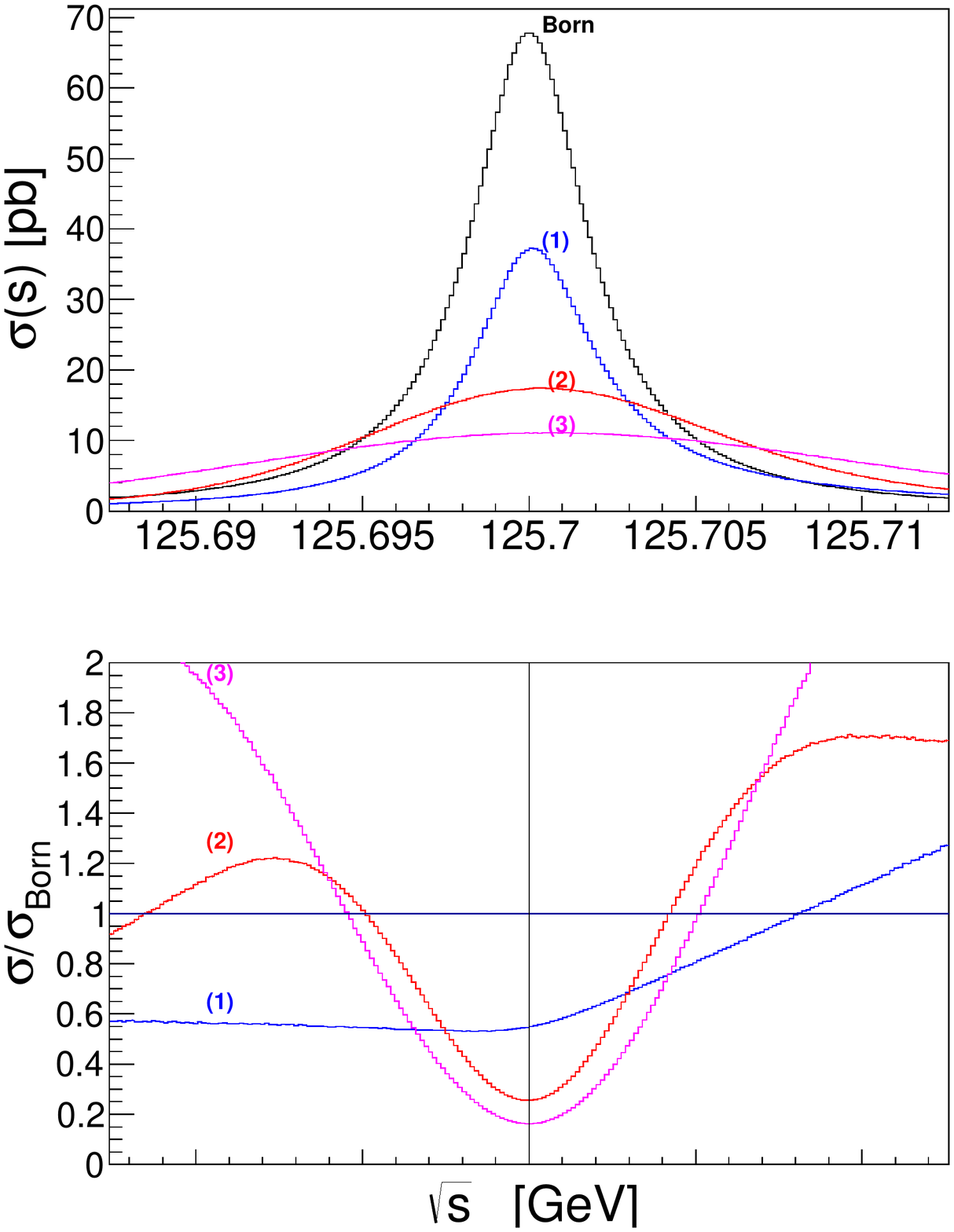}
  \caption{\sf
    Line-shape of the direct Higgs production process in the $\mu^+\mu^-$ collider
    including effect of ISR type and the machine energy spread.
    The same input and notation as in Fig.~\ref{fig:Comparison}
  }
  \label{fig:muonComparison}
\end{figure}

Numerical results for muon collider are presented in
Fig.~\ref{fig:muonComparison} and Tab.~\ref{tab:muonComparisioDistributionsPeakALL}, 
and correspond to Higgs production cross section dependence on energy $E$ and machine
energy spread $\delta$ shown in Fig.~\ref{fig:Comparison} and 
in Tab.~\ref{tab:ComparisioDistributionsPeakALL}.
As seen in this table, reduction factor of the cross section
at the resonance peak due to QED ISR is now only 0.548.
For the machine energy spread $\delta=8$MeV it deteriorates down to 0.162
(which is accidentally comparable to electron collider case with $\delta=4$MeV)
and for $\delta=4.2$MeV it is equal 0.279%
\footnote{This agrees only rather crudely with 0.25 result
 in ref.~\cite{Greco:2015yra} for $\delta=4$MeV, obtained
 from approximate analytical formula.}.

\section{Summary}

It was analysed numerically what is the influence of
the QED ISR and machine energy spread on the 
the resonant Higgs boson production cross section,
the so called Higgs line-shape,
in the process $e^+e^-\rightarrow H$ and $\mu^+\mu^-\rightarrow H$.

It was found that for electron collider
the QED ISR reduces by itself peak cross section by factor 0.338.
The QED higher order uncertainty of this results
was estimated to be below $0.3\%$.
The proper double convolution of the QED radiative spectrum with the
machine energy spread was performed and doubly cross-checked.
For instance machine energy spread of the same size as Higgs width
($\sigene\simeq 4$MeV) reduces Higgs production
cross section further down to 0.172$\sigma_{B}$.
These results are compatible (albeit slightly different) from
these in the preliminary analysis shown at the recent FCCee workshop,
see \cite{8th_FCC_Workshop_presentation}.
It forms a solid basis for any analysis of the observability
of the Higgs resonant cross section in the future $e^+e^-$ colliders.

The same analysis was also repeated for the muon colliders,
where the reduction factor due to QED ISR at the peak position was found
to be 0.548 and, for instance,  it gets reduced further down to 0.162 
for the energy beam spread being twice the Higgs width.

\newpage
\appendix
\section{Non-uniqueness of the Breit-Wigner formula}
\label{Appendix:MassShift}

In this short Appendix we want to comment on the issue of the 
form of the non-uniqueness of the Breit-Wigner formula (\ref{Breit-Wigner})
and we shall show that for extremely narrow Higgs boson resonance
this non-uniqueness is numerically completely irrelevant.

The literature on this question is numerous, for instance in the context
of the precision measurements of the $Z$ and $W$ boson production and decay
the reader may consult 
refs.~\cite{Bardin:1988xt,Stuart:1991xk,Stuart:1991cc,Dittmaier:2012vm}.
In particular the use of 
the complex energy poles of the propagator of virtual particles as a natural 
solution was advocated, see ref.~\cite{Dittmaier:2012vm}.

Following ref.~\cite{Bardin:1988xt} (eqs. (1.6) through (1.9)),
the Born term corresponding to (\ref{Breit-Wigner}) is 
\begin{equation}
 \sigma_{B}(s)\sim \frac{s}{(s-M_{H}^{2})^{2}+G^{2}}.
 \label{discussion_Born}
\end{equation}
In case of constant width $G=M_{H}^{2}\gamma_H = 527.94$
where $ \gamma_H=\frac{\Gamma_{H}}{M_{H}}= 3.34\cdot 10^{-5} $,
while for the $s$ dependent one we have $G=G(s)=s\gamma_H$. 
The maximum of the cross section is at
\begin{equation}
 \sqrt{s_0}=E_{0}=M_{H}(1+\gamma_H^{2})^{1/4}
 \simeq M_H \big(1+\frac{\gamma_H^{2}}{4}\big)
 =125.700000035~{\rm GeV}
\end{equation}
for the fixed width and 
\begin{equation}
 E_{0}=M_{H}(1+\gamma_H^{2})^{-1/4}
 \simeq M_H \big(1-\frac{\gamma_H^{2}}{4} \big)
 =125.69999996~{\rm GeV}
\end{equation}
for $s$-dependent width.
The difference between $E_0$ in the above two cases
may be regarded as $7 \times 10^{-8}~$GeV. To our knowledge it 
is practically impossible to achieve energy resolution in accelerators 
of that order of magnitude.

Finally, the formula with the so-called running width 
(for different variants of Breit-Wigner see, e.g., \cite{Anastasiou:2012hx} 
and references therein) reads
\begin{equation}
 \sigma_{B}(s) \sim \frac{1}{(s-M_{H}^{2})^{2}+G^{2}},
\end{equation}
where $G=s\gamma_{H}$.
The maximum of the cross section is at
\begin{equation}
 \sqrt{s_{H}}=\frac{M_{H}}
 {\sqrt{1+\gamma_{H}^{2}}}
 \simeq M_H \big(1-\frac{\gamma_H^2}{2} \big),
\end{equation}
which, as above, is beyond measurability in any future accelerator experiment.

\section{Approximate formulas for QED ISR}

Starting from eq.~(\ref{sigmaISR}) we are going to provide
a very simple approximate formula that allows to estimate the suppression
factor of the peak cross section at $s=M_{H}^{2}$ due to QED ISR corrections,
thus providing quick/easy cross-check of
more sophisticated numerical calculations.

In the most simplified soft photon approximation ISR 
formula of eq.~(\ref{sigmaISR}) at $E=M_{H}$ reads
\begin{equation}
 \sigma(M_{H})\approx \int_{0}^{1}\gamma v^{\gamma-1} \sigma_{B}
 (M_{H}^{2}(1-v)) dv.
\end{equation}

It is well known fact that Breit-Wigner profile drop sharply
around $|s-M_H^2|=\Gamma_H^2$ and one may therefore approximate
it by the following rectangular shape
\[
\sigma_B(s) \simeq \sigma_B(M_H^2) \theta( |s-M_H^2|<\Gamma_H^2),
\]
which translates into $v \leq \Gamma_H/M_H$ integration limit.
As a result we 
obtain the following approximate suppression factor for ISR corrections
\begin{equation}
r_{ISR}=
\frac{\sigma_{I}(M_{H}^{2})}{\sigma_{B}(M_{H}^{2})} 
\simeq \frac{1}{\sigma_{B}(M_{H}^{2})}
\int_{0}^{\frac{\Gamma_{H}}{M_{H}}}\gamma v^{\gamma-1} \sigma_{B}
(M_{H}^{2})  dv = \left(\frac{\Gamma_{H}}{M_{H}}\right)^\gamma
\end{equation}
For $e^+e^-$ colliders  with $\gamma =  0.1106$
the ISR suppression factor is then estimated to be
\begin{equation}
 r_{ISR} =  0.3199.
\end{equation}
This agrees reasonably well with the exact ISR factor of $0.347$,
seen in Tab.~\ref{tab:ComparisioDistributionsPeak}. 
The relative error of the approximate formula is about $8\%$
\footnote{ Sometimes one expands further
 $ r_{ISR} \simeq 1- \gamma \ln(\Gamma_H/M_H) $
 but this makes sense only for $r_{ISR} > 0.5$, which is not our case.}.
For $\mu^+ \mu^-$ collider with smaller $\gamma = 0.0611 $ approximate
ISR suppression factor
\begin{equation}
 r_{ISR} =  0.5329, 
\end{equation}
is obtained,
to be compared with the corresponding value of $0.548$
in Tab.~\ref{tab:muonComparisioDistributionsPeakALL}. 
The relative error of the approximate formula is now merely $3\%$.

\end{document}